\documentclass[preprint,letter,preprintnumbers,amsmath,amssymb,floatfix]{revtex4}

\usepackage{graphicx}
\usepackage{dcolumn}
\usepackage{bm}

\def\lsim{\mbox{~{\protect\raisebox{0.4ex}{$<$}}\hspace{-1.1em}
        {\protect\raisebox{-0.6ex}{$\sim$}}~}}

\begin{document}

\title{Dilepton production from a viscous QGP}
\author{Kevin Dusling and Shu Lin}
\affiliation{Department of Physics \& Astronomy, State University of New York, Stony Brook, NY 11794-3800, U.S.A.}
\date{\today}  

\begin{abstract}

This work calculates the first correction to the leading order q\={q} dilepton production rates due to shear viscosity in an expanding gas.  The modified rates are integrated over the space-time history of a viscous hydrodynamic simulation of RHIC collisions.  The net result is a {\em hardening} of $q_\perp$ spectrum with the magnitude of the correction increasing with invariant mass.  We argue that a thermal description is reliable for invariant masses less than $M_{max}\approx(2\tau_0 T_0^2)/(\eta/s)$.  For reasonable values of the shear viscosity and thermalization time $M_{max}\approx 4.5$ GeV.  Finally, the early emission from a viscous medium is compared to emission from a longitudinally free streaming plasma.  Qualitative differences in $q_\perp$ spectrum are seen which could be used to extract information on the thermalization time, viscosity to entropy ratio and possibly the thermalization mechanism in heavy-ion collisions.        
 
\end{abstract}

\maketitle  

\section{Introduction}

There is a general consensus that the early matter produced at RHIC behaves as a near perfect fluid \cite{Ollitrault:2006va}.  This conclusion was born out of the success of ideal hydrodynamic descriptions \cite{Teaney:2001av,Kolb:2003dz} of both hadron transverse momentum spectra and elliptic flow measurements up to 1.5-2 GeV/c.  Although it is too early to draw any definitive conclusions most likely the deviations from ideal hydrodynamic behavior can be ascribed to dissipative effects. This has already been suggested in some of the recent works on dissipative hydrodynamics \cite{Dusling:2007gi,Song:2007ux,Baier:2006gy,Romatschke:2007jx,Bozek:2007qt,Song:2007fn,Chaudhuri:2007qp}.

In addition to hadronic observables which interact strongly and therefore depend only on the final state of the medium, electromagnetic probes are emitted throughout the entire space-time evolution reaching the detector without any final state interactions.  In terms of computing observables there is a big difference, since the resulting transverse momentum and elliptic flow spectra depend only on the final freezeout hypersurface, whereas the resulting dilepton yields depend on the full space-time volume.  A consistent description of heavy-ion phenomenology should use the same space-time evolution for both hadronic spectra and dilepton observables.

In this work we calculate the first viscous correction to dilepton emission from quark anti-quark annihilation in a dissipative medium.  The kinematic region when a thermal description is reliable is found by requiring that the viscous corrections are small.  When the viscous corrections become large a kinetic description is really required.  The viscous rates are then integrated over the space-time history of a hydrodynamic simulation of RHIC collisions.  We show how shear viscosity modifies the transverse momentum and invariant mass spectrum.  We find that the inverse slope of the transverse mass spectrum is sensitive to both the thermalization time as well as the shear viscosity and can therefore be used in order to learn about the early stages of heavy-ion collisions.  Finally a comparison is made with dileptons produced from a free-streaming quark-gluon plasma.   

\section{Dilepton Rates}

The rate of dilepton emission from a quark-gluon plasma due to q\={q} annihilation is given in the Born approximation as

\begin{equation}
\frac{dN}{d^4q}=\int\frac{d^3{\bf k}_1}{(2\pi)^3}\frac{d^3{\bf k}_2}{(2\pi)^3} f(E_1,T) f(E_2,T) v_{12} \sigma(M^2) \delta^4(q-k_1-k_2)\,,
\label{eq:KT}
\end{equation}

where $q=(q_0,{\bf q})$ is the virtual photon's four momentum and $M^2=(E_1+E_2)^2-({\bf k}_1 +{\bf k}_2)^2$ is the photon's invariant mass.  Throughout this work we consider massless quarks; therefore $E_{1,2}=\sqrt{{\bf k}_{1,2}^2+m_q^2}\approx|{\bf k}_{1,2}|$. The function $f(E,T)$ is the quark or anti-quark momentum distribution function, which in thermal equilibrium is given by $f(E,T)=1/(1+e^{E/T})$.

In the above expression $v_{12}$ is the relative velocity of a quark anti-quark pair and $\sigma(M^2)$ is the q\={q} cross section.  Both expressions are well known from the literature \cite{WongBk} and are given by $v_{12}=\frac{M^2}{2 E_1 E_2}$ and  $\sigma(M^2)=\frac{16\pi\alpha^2 e_q^2N_c}{3M^2}$.  The integral over the quarks' momentum can be done analytically with the result

\begin{equation}
\frac{dN}{d^4q} = -\frac{\alpha^2}{12 \pi^4} (N_c \sum_{u,d,s}e_q^2) f_b(q_0,T)\left[1+\frac{2T}{|{\bf q}|}\ln(\frac{n_+}{n_-})\right]\,,
\end{equation}

where $n_\pm=1/(e^{(q_0\pm|{\bf q}|)/2T}+1)$ and $f_b(q_0,T)=1/(e^{q_0/T}-1)$.

\section{Viscous Correction to the Dilepton Rates}

In order to account for dissipative effects in the dilepton emission rate we include the first viscous correction to the quark and anti-quark's distribution function in eq. \ref{eq:KT}.  This approach neglects any space-time inhomogeneities and assumes that the distribution functions relax to their dissipative forms much quicker than the medium evolves.  Ideally, one could solve the Baym-Kadanoff equations which would take non-equilibrium effects into account.  We note that the leading order born q\={q} rates do not contain pinch singularities which suggests that at least to leading order one may neglect space-time inhomogeneities \cite{Gelis:2001xt}.  This approximation allows us to calculate the dilepton emission rates locally in a space-time volume $d^4x$. 

As shown in \cite{Teaney:2003kp,Arnold:2000dr,GrootBk} viscosity modifies the ideal distribution function.  The resulting correction for fermions is

\begin{equation}
f(k)\to f(k)+\frac{C_1}{2 (\epsilon + p) T^2}f(k)[1-f(k)]k^\alpha k^\beta\pi_{\alpha\beta}\,,
\label{eq:vdist}
\end{equation}

where $\pi_{\alpha\beta}=\eta\langle \nabla_\alpha u_\beta \rangle\equiv\eta( 
\nabla_\alpha u_\beta+\nabla_\beta u_\alpha-\frac{2}{3}\Delta_{\alpha\beta}\nabla_\rho u^\rho)$ and $\eta$ is the shear viscosity not be confused with the space-time rapidity $\eta_s$.  The coefficient $C_1$ can be computed analytically for a massless fermion gas and is given by $C_1=14\pi^4/1350\zeta(5)\approx0.97$. Substituting the above result into the born annihilation rate (eq.~\ref{eq:KT}) and keeping terms up to first order in 
$\eta/s$ (quadratic in momentum) one obtains:

\begin{eqnarray}
\frac{dN}{d^4q}&=&\frac{4 N_c\alpha^2 e_q^2}{3(2\pi)^5}\int \frac{d^3{\bf k}_1 d^3{\bf k}_2}{E_1E_2}\delta^4(q-k_1-k_2) \times \nonumber\\
&&\left[ f(k_1)f(k_2) + \left(\frac{C_1}{2 (\epsilon + p) T^2}f(k_1)[1-f(k_1)]f(k_2)k_1^\alpha k_1^\beta+k_1\leftrightarrow k_2\right)\pi_{\alpha\beta}\right] \nonumber\\
&=&\frac{4 N_c\alpha^2 e_q^2}{3(2\pi)^5}\int \frac{d^3{\bf k}_1 d^3{\bf k}_2}{E_1E_2}\delta^4(q-k_1-k_2)\times\nonumber\\
&&\left[ f(k_1)f(k_2)+\frac{C_1}{(\epsilon+p) T^2}f(k_1)[1-f(k_1)]f(k_2) k_1^\alpha k_1^\beta\pi_{\alpha\beta}\right]
\label{eq:KTv1}
\end{eqnarray}

In simplifying the above result we have used the fact that the permutation of $k_1\leftrightarrow k_2$ gives the same contribution after integration.  We write the final result as the sum of the ideal and viscous correction

\begin{equation}
\frac{dN}{d^4q}= I_1(q) + \frac{C_1}{(\epsilon+p)T^2}I_2^{\alpha\beta}(q)\pi_{\alpha\beta}\,,
\end{equation}

with

\begin{eqnarray}
I_1&=&-\frac{N_c\alpha^2 e_q^2}{12 \pi^4}f_b(q_0)\left[1+\frac{2T}{|{\bf q}|}\ln(\frac{n_+}{n_-})\right]\,,\nonumber\\
I_2^{\alpha\beta}&=&\frac{4 N_c\alpha^2 e_q^2}{3 (2\pi)^5}\int\frac{d^3{\bf k}_1}{E_1 E_2}f(E_1)[1-f(E_1)]f(E_2)k_1^\alpha k_1^\beta\delta(E_1+E_2-q_0)\,.
\end{eqnarray}

Since $I_2^{\alpha\beta}$ is a second rank tensor depending only on $u^\alpha$ and $q^\alpha$ it can be decomposed as

\begin{equation}
I_2^{\alpha\beta}=a_0 g^{\alpha\beta}+a_1 u^\alpha u^\beta+a_2 q^\alpha q^\beta + a_3(u^\alpha q^\beta + u^\beta q^\alpha)\,.
\end{equation}

The final result will contain the term $I_2^{\alpha\beta}\pi_{\alpha\beta}$.  By making use of the identities $u^\alpha\pi_{\alpha\beta}=g^{\alpha\beta}\pi_{\alpha\beta}=0$ only the term with coefficient $a_2$ will be non-vanishing.  $a_2$ is found by using the identity $a_2=P_{\alpha\beta}I_2^{\alpha\beta}$ where the projection operator in the local rest frame of the medium is

\begin{equation}
P_{\alpha\beta}=\frac{1}{2|{\bf q}|^4}[(3q_0^2-|{\bf q}|^2)u_{\alpha}u_{\beta}
-6q_0u_{\alpha}q_{\beta}+3q_{\alpha}q_{\beta}+|{\bf q}|^2g_{\alpha\beta}]\,.
\end{equation}

We now quote the final result for the first viscous correction to the born dilepton annihilation rates:

\begin{equation}
\frac{dN}{d^4q}=-\frac{N_c\alpha^2 e_q^2}{12\pi^4}\left[f_b(q_0,T)[1+\frac{2T}{|{\bf q}|}\ln(\frac{n_+}{n_-})]-\frac{C_1}{2(\epsilon+p)T^2} b_2(q_0,|{\bf q}|)q^\alpha q^\beta\pi_{\alpha\beta}\right]
\label{eq:KTvis}
\end{equation}

where we have defined

\begin{equation}
b_2(q_0,|{\bf q}|)= \frac{1}{|{\bf q}|^5}\int_{E_-}^{E_+}f(E_1,T)f(q_0-E_1)(1-f(E_1))\left[ (3q_0^2-|{\bf q}|^2)E_1^2-3q_0E_1M^2+\frac{3}{4}M^4\right]dE_1
\label{eq:b2}
\end{equation}

and $E_{\pm}=\frac{1}{2}(q_0\pm|{\bf q}|)$.  For large invariant masses $(M/T \gg 1)$ one can replace the Fermi distribution with the classical Maxwell-Boltzmann distribution.  In the viscous correction to the distribution function this amounts to substituting $f_f(1-f_f)\to f_{MB}$.  In this limit an analytic expression can be found for the viscous correction to the dilepton rates.  In the limit that $(u\cdot q)/T \gg 1$ the resulting expression is given as

\begin{equation}
\frac{dN}{d^4q}=\frac{N_c\alpha^2 e_q^2}{12\pi^4}e^{-q_0/T}\left[1+\frac{C_1}{3(\epsilon+p)T^2}q^\alpha q^\beta\pi_{\alpha\beta}\right]\,,
\label{eq:appx}
\end{equation}

where as before $C_1\approx 0.97$.  We find that the above result holds at the accuracy of a few percent for $M \geq 3$ GeV at $T=400$ MeV.

A feature of the viscous correction is that it does not modify the invariant mass spectrum.  This is seen by looking at either of the above forms of the viscous correction (eq.~\ref{eq:appx} or \ref{eq:KTvis}).  In going from $d^4q$ to $dM^2$ the resulting integral will be a second rank tensor depending on $u^\alpha$ only.  The most general form the result can take is a linear combination of terms proportional to $g^{\alpha\beta}$ and $u^\alpha u^\beta$, which both vanish when contracted with $\pi_{\alpha\beta}$.  

Before performing the full space-time evolution we illustrate the effect of the viscous correction using
a simple model for the gradients.  We consider a 1D Bjorken expansion without transverse flow.  The viscous component of the stress-energy tensor can be easily computed and is given as 

\begin{equation} 
q^{\alpha}q^{\beta}\langle\nabla_\alpha u_\beta\rangle=\frac{2}{3\tau}q_{\perp}^2-\frac{4}{3\tau}m_{\perp}^2 \sinh^2(y-\eta_s)\,.
\label{eq:stressBj}
\end{equation}

\begin{figure}
\includegraphics[width=0.45\textwidth]{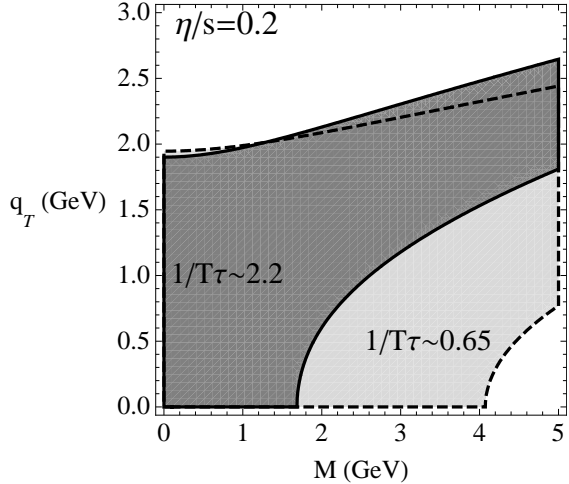}
\caption{\label{fig:map} Kinematic regions where the viscous correction is less then order one.  More precisely, the boundary is set by the condition $\lvert \delta f/f_0\rvert \leq 0.8$.}
\end{figure}

By substituting the above result into eq. \ref{eq:appx} an analytic expression can be found for the dilepton yields in the limit that $M/T\gg 1$.  After performing the integration over $\eta_s$ the result is

\begin{equation}
\frac{dN}{dM^2 dq_\perp^2 dy}=\frac{N_c\alpha^2 e_q^2}{12\pi^3}K_0(x)\left(1+\frac{2 C_1}{9\tau T}\left(\frac{\eta}{s}\right)\left[\left(\frac{q_\perp}{T}\right)^2-2\left(\frac{m_\perp}{T}\right) \frac{K_1(x)}{K_0(x)}\right]\right)\,,
\label{eq:Be}
\end{equation}

where $K_\nu(x)$ is the modified Bessel function evaluated at $x\equiv m_\perp/T$.  Fig.~\ref{fig:map} shows the kinematic regions where the viscous correction is small ({\em i.e.} $dN_{vis}/dN_{ideal} \leq 0.8$) and therefore dictates when using a thermal description of dilepton production is suitable.  The criterion that dictates when hydrodynamics is applicable can be written as

\begin{equation}
\left(\frac{\eta}{s}\right)\times\frac{1}{\tau T} \ll 1
\end{equation}

and can therefore be separated into a condition on the medium, $\eta/s$, and a condition on the experimental setup, $1/(\tau T)$.  Throughout this work we always set $\eta/s=0.2$.  The region surrounded by the solid line is for $1/(\tau T)\approx 2.2$ corresponding to a temperature of 450 MeV at $\tau=0.2$ fm/c.  The region surrounded by the dotted line is for $1/(\tau T)\approx 0.65$ corresponding to a temperature of 300 MeV at $\tau=1$ fm/c.  At earlier times the viscous correction is larger and the allowed region is smaller.

The results shown in fig.~\ref{fig:map} should only be taken qualitatively.  Transverse flow alleviates the situation, opening up the boundaries shown above.  Also, the result presented was in the limit $M/T >> 1$ where an analytical result was obtained.  Fig.~\ref{fig:map} is still useful, since it still serves as a qualitative picture where the viscous corrections become large even after including flow and relaxing the classical limit.  Outside of the kinematic boundaries a thermal description may no longer be reliable.

\begin{figure}
\includegraphics[width=0.45\textwidth]{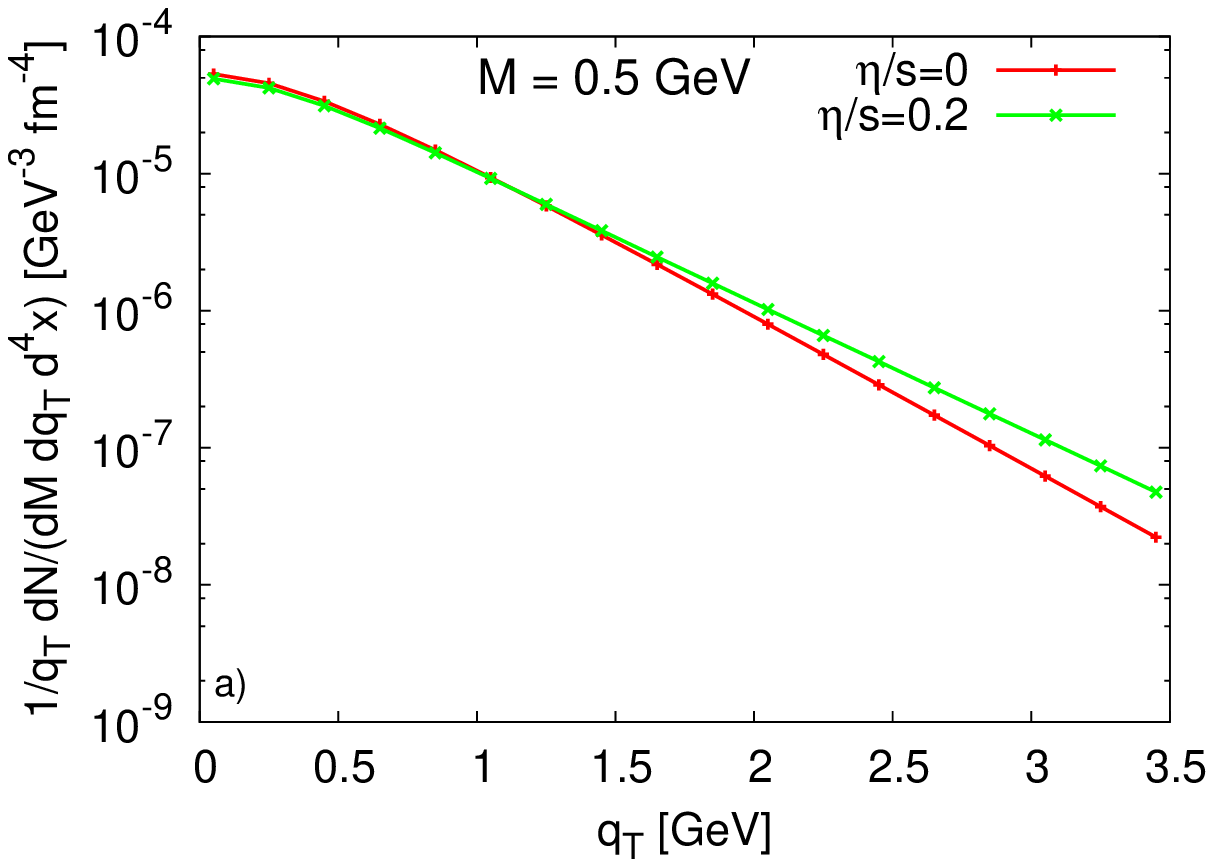}
\includegraphics[width=0.45\textwidth]{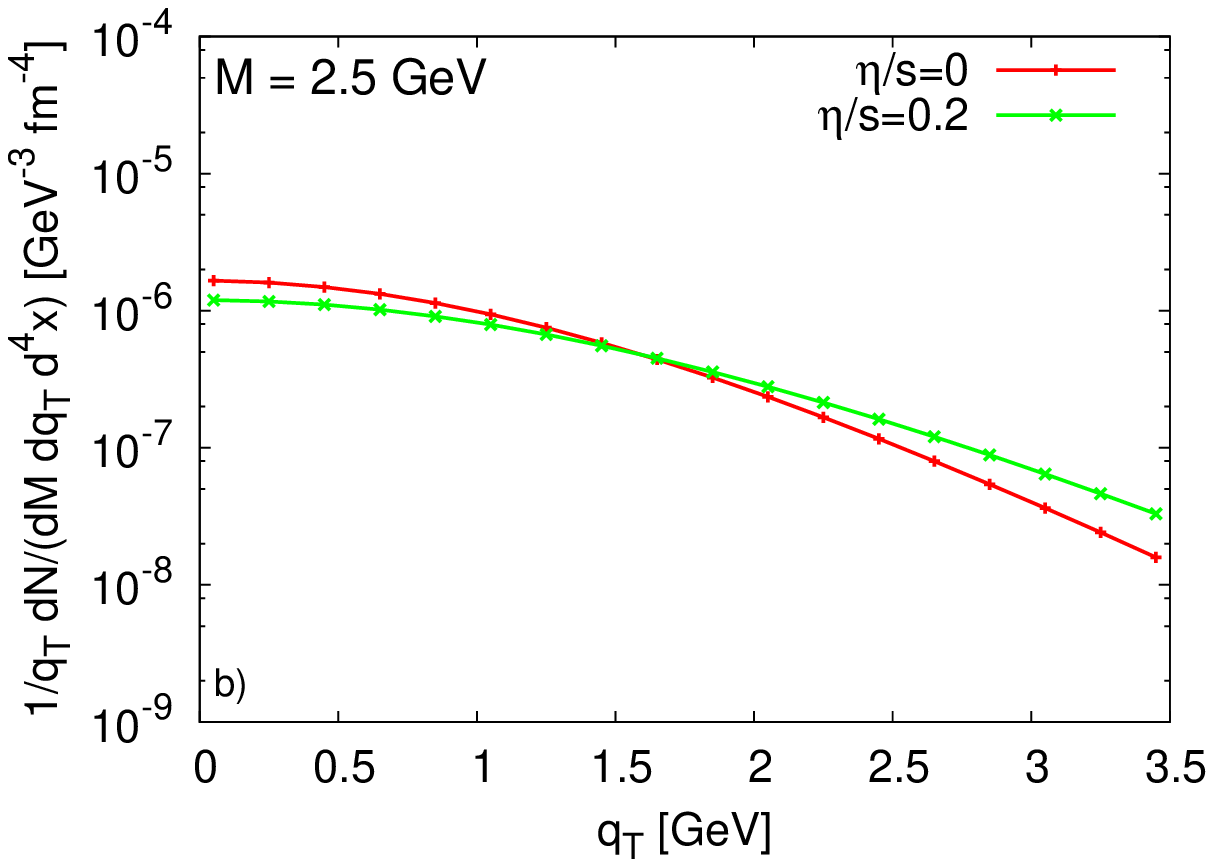}
\caption{\label{fig:qt}(color online) Dilepton transverse momentum spectra
 for $T=0.4$ GeV and $\eta/s=0.2$ at $\tau=1$ fm/c for a boost invariant expansion with no transverse flow.}
\end{figure}

Throughout the remainder of this work we now resort to eq.~\ref{eq:KTvis}, which is evaluated numerically, in order to compute the dilepton yields accurately at all masses.  Fig.~\ref{fig:qt} shows the dilepton spectrum generated for a temperature T=0.4 GeV at proper time $\tau=1$ fm/c and using a viscosity to entropy ratio of $\eta/s=0.2$. 

Figures of the invariant mass spectrum are not shown because, as discussed earlier, the spectrum is unmodified when including the viscous correction.  Looking at the transverse momentum spectrum a {\em hardening} of dileptons is seen that is reminiscent of the single particle spectrum in \cite{Teaney:2003kp}.  The magnitude of the viscous correction is dictated by both $\eta/s$ as well as the proper time.  At earlier times the shear between the longitudinal and transverse directions is larger resulting in bigger corrections at smaller proper times due to the $1/\tau$ factor in eq.~\ref{eq:stressBj}.

\section{Evolution Model}

In order to model the space time evolution of the collision region we use the results of \cite{Dusling:2007gi}, which is summarized in this section.  The hydrodynamic model is a 1+1 dimensional boost invariant expansion with initial conditions tuned in order to simulate Au-Au collisions at RHIC energies ($\sqrt{s}=200$ GeV).  Dissipative corrections to the ideal hydrodynamic expansion is treated using a second order relaxation scheme first proposed by \"{O}ttinger and Grmela \cite{OG}.

Detailed plots of the hydrodynamic solution with and without viscosity is shown in \cite{Dusling:2007gi}.  For modest values of the shear viscosity ($\eta/s \lsim 0.3$) dissipative effects did not integrate to give large changes to the ideal hydrodynamic solution.  
The net effect of a finite viscosity was twofold.  First, the longitudinal pressure is initially reduced causing a slower decrease of energy density per unit rapidity at early times.  The reduction of longitudinal pressure is accompanied by a larger transverse pressure which drives larger transverse velocities.  The larger velocities then cause the energy density to deplete faster at later times.

Even though the changes to the ideal hydrodynamic result is small a full viscous simulation is still needed in order to have access to the velocity gradients which enter into the dissipative corrections of the quark and anti-quark distribution functions.

The hydrodynamic model is started at $\tau_0=0.2$ fm/c in order to account for some of the pre-equilibrium production of dileptons which will contribute at larger masses.  Dileptons are produced as long as the temperature of the medium is greater than a critical temperature taken as $T_c=0.170$ GeV.  We do not look at dileptons produced during a mixed phase or hadronic phase in this work.  At any space-time point the hydrodynamic model provides the three independent terms of the stress tensor ($\pi^{rr}, r^2\pi^{\phi\phi}$ and $\tau^2\pi^{\eta\eta}$).  The equation for $q^{\alpha}q^{\beta}\pi_{\alpha\beta}$ used in eq.~\ref{eq:KTvis} is given as   

\begin{eqnarray}
q^{\alpha}q^{\beta}\pi_{\alpha\beta}&=&q_{\perp}^2\cos^2(\theta)\pi^{rr} +q_{\perp}^2\sin^2(\theta)r^2\pi^{\phi\phi} + m_{\perp}^2\sinh^2(\eta_s) \tau^2\pi^{\eta\eta} \nonumber\\ && + m_{\perp}^2\cosh^2(\eta_s) v^2\pi^{rr} - 2m_{\perp}\cosh(\eta_s) q_{\perp}\cos(\theta)v\pi^{rr}\,.
\end{eqnarray}

\begin{figure}
\includegraphics[width=0.45\textwidth]{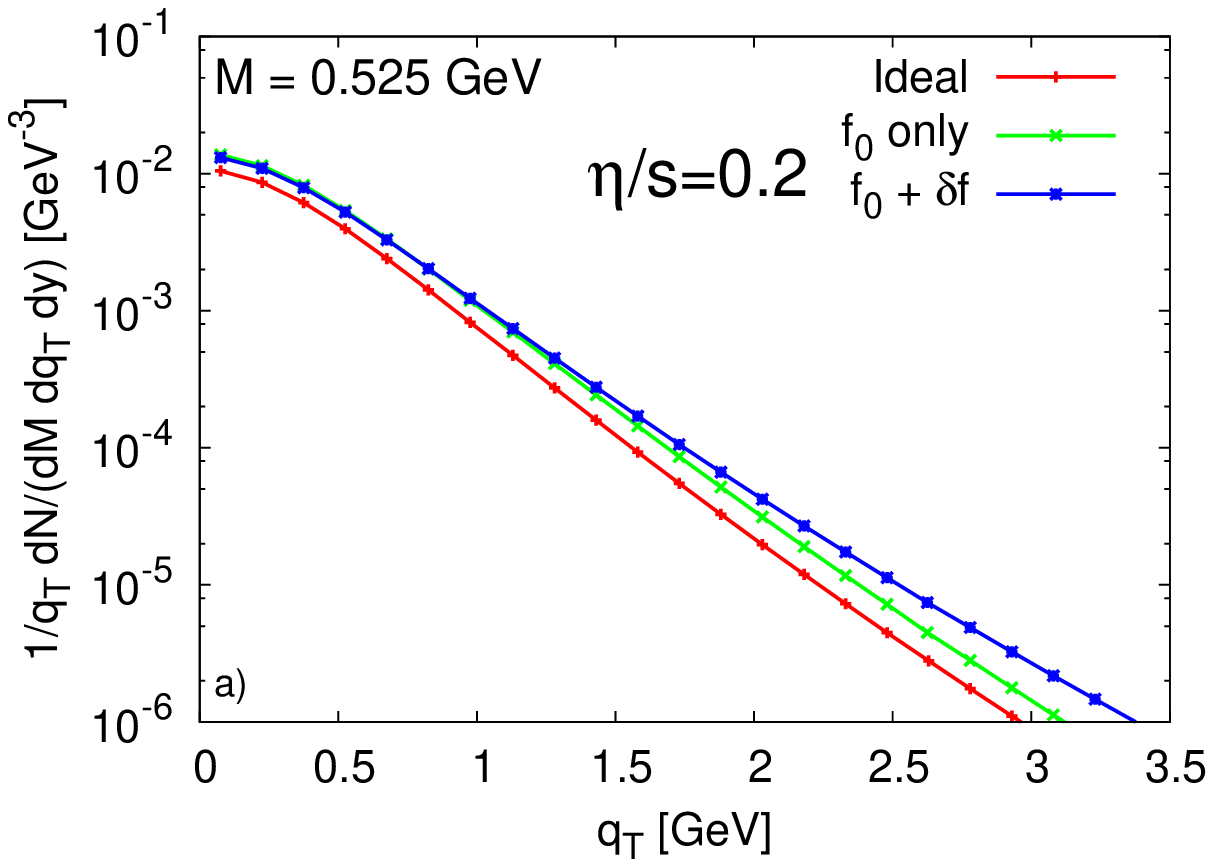}
\includegraphics[width=0.45\textwidth]{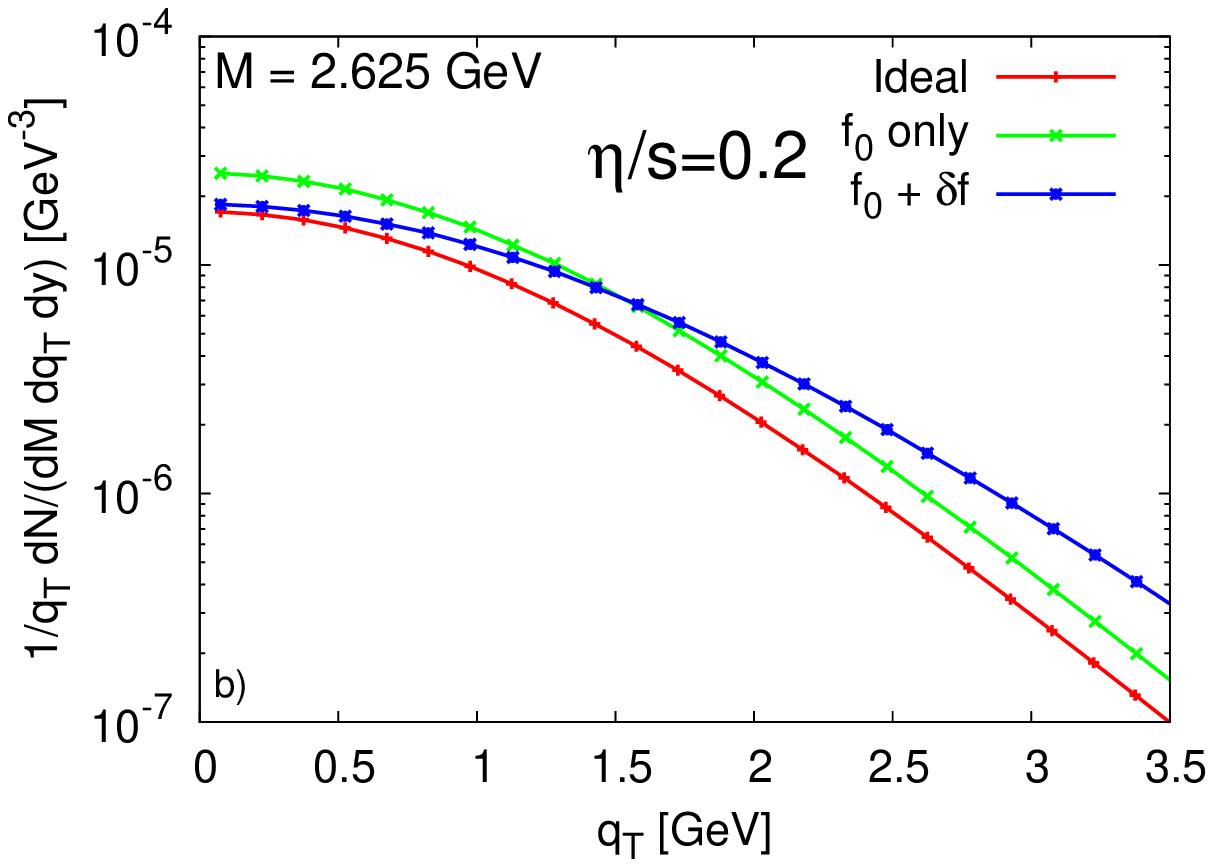}
\caption{\label{fig:qtINT}(color online) Dilepton transverse momentum spectra
 after the full space-time integration of a boost invariant expansion with arbitrary transverse expansion and azimuthal symmetry.}
\end{figure}

Fig.~\ref{fig:qtINT} shows the resulting transverse momentum spectrum after the full space-time integration at two invariant mass points: $M=0.525$ GeV (left) and $M=2.625$ GeV (right).  First the red curve shows spectra generated from an ideal hydrodynamic simulation ($\eta/s=0$).  Next the green curve shows the spectra generated from a viscous simulation having $\eta/s=0.2$ but without including the viscous correction to the distribution function.  This curve therefore shows the effect that viscosity has on modifying the ideal hydrodynamic equation of motions.  We find that a finite viscosity leads to a slight increase in the overall yield.  This is due to the fact that a finite viscosity causes the energy density to deplete more slowly at early times.  This effect therefore brings about an effective increase in the lifetime of the simulation above the critical temperature. We find $\approx 30\%$ increase in the low mass region and $\approx 50\%$ increase in the higher mass region.  

Finally, the blue curves in fig. \ref{fig:qtINT} show the viscous result including the viscous correction to the distribution function.  We find that the magnitude of the viscous correction increases with the invariant mass.  This was similarly observed in fig.~\ref{fig:map} where the range in $q_\perp$ having viscous corrections of order less than one (as shown by shaded regions) decreased in size with increasing mass.  The simulation results are discussed in more detail in the next section.    

\section{Discussion}

In order to further understand the viscous corrections, the effective temperature ($T_{eff}$) of the dilepton spectrum from the full space-time integration is constructed as a function of invariant mass.  The effective temperature is found by fitting the transverse mass spectrum at a given mass to the following expression, 

\begin{equation}
\frac{dN}{dM^2 m_\perp dm_\perp dy} \propto e^{-m_\perp/T_{eff}}\,.
\end{equation}

In this work the fit is done in the transverse momentum region of $0.5\leq q_\perp \text{(GeV)} \leq 2.0$.  As expected, we find that the transverse mass spectra does not exactly fit the above form.  Actually, other ranges in $q_\perp$ could have been chosen where the fit works better.  However, the results are qualitatively the same and therefore a different choice in $q_\perp$ range will not change the discussion that follows.  If a quantitative comparison were to be made with data, it would be more appropriate to compare to the actual $q_\perp$ spectra instead.  Regardless, $T_{eff}$ still serves as a useful quantity since it probes the average temperature of the medium as well as the radial flow profile and viscous correction.  Looking at fig.~\ref{fig:qtINT} we expect the viscous correction to increase the effective temperature, with larger corrections at higher masses.

\begin{figure}
\includegraphics[width=0.6\textwidth]{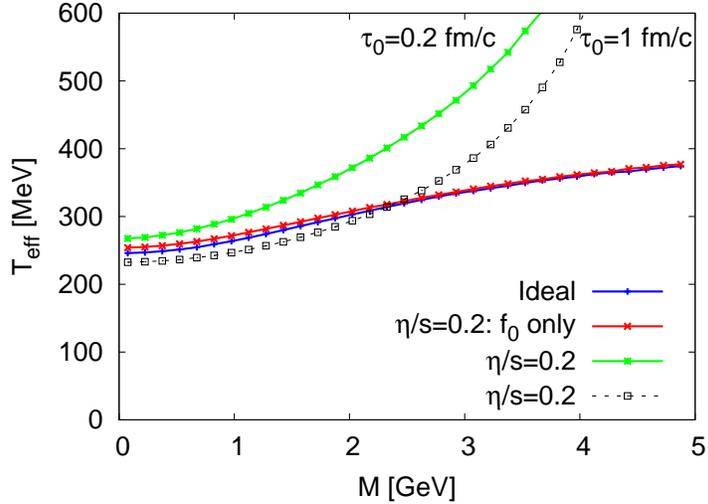}
\caption{\label{fig:Teff}(color online) Effective temperature as a function of invariant mass.}
\end{figure}

In fig.~\ref{fig:Teff} the effective temperature is shown as a function of invariant mass.  The solid blue curve labeled {\em ideal} shows $T_{eff}$ for an ideal ($\eta/s=0$) hydrodynamic expansion started at proper time $\tau_0=0.2$ fm/c.  The shape of the curve is dictated by the underlying radial flow as well as the average temperature of the emission region.  At higher masses the dominate source of dileptons is from the higher temperature regions, which occur at earlier proper times.  This explains the slight rise in $T_{eff}$ with mass.  We now look at the solid red curve, which is generated from a hydrodynamic evolution having $\eta/s=0.2$.  In this case we do not include the viscous correction to the distribution function and the resulting modifications to the effective temperature are due to changes in the hydrodynamic evolution.  As discussed earlier, modifications from viscosity to the hydrodynamic solution are small and we therefore don't expect to see large deviations from the ideal case.  This is indeed the case.

We now focus the discussion on the role of the viscous correction to the distribution function.  The green curve in fig.~\ref{fig:Teff} shows the effective temperature of dileptons coming from a viscous medium having $\eta/s=0.2$ from a simulation started at a proper time of $\tau_0=0.2$ fm/c.  The result is that the effective temperature increases greatly as a function of invariant mass.  From the magnitude of the correction, the upper bound of the domain of hydrodynamics is found to be at most $M_{max}\approx 2.0$ GeV for this parameter set.  There are two reasons why the viscous correction increases with mass.  First, there is the explicit mass dependence in the viscous correction itself.  This is easiest to see by looking at the approximate form, eq.~\ref{eq:Be}.  The second reason is because the high mass contribution is mainly produced in the early, high temperature stages of the evolution.  Looking at eq.~\ref{eq:stressBj} the viscous correction grows like $1/\tau$ at early times.  In order to see the effect of the early emission a final simulation is done (dashed-black curve) where the hydrodynamic evolution is started at $\tau_0=1$ fm/c.  In this case the viscous corrections are more modest and $M_{max}\approx 4.5$ GeV.          

It is therefore a combination of both the thermalization time as well as the magnitude of $\eta/s$ that dictates when a hydrodynamic description is reliable.  Since the effective temperature rises so quickly with mass, as long as there is non-vanishing viscosity, there will always be a mass region outside of the region of a hydrodynamic description.  From eq.~\ref{eq:Be} one can extract an approximate condition for the mass.  Since most of the particle yield is at low $q_\perp$ we should guarantee that the viscous correction is small at $q_\perp=0$.  Furthermore at high and intermediate masses the ratio of Bessel functions is approximately one.  Then the condition that the viscous correction must be less than of order one can be expressed as

\begin{equation}
M_{max} \approx \frac{2 \tau_0 T_0^2}{\eta/s}\,.
\end{equation}

When the viscous corrections to the spectra become large a kinetic approach is required.  One can ask whether the viscous correction at early times {\em mock up} the effects of off-equilibrium production that would be taken into account by a full kinetic theory.  In order to test this hypothesis $T_{eff}$ spectra is calculated from a free streaming non-interacting gas of quarks \cite{Gyulassy:1997ib,Kapusta:1992uy}.  We should point out that our treatment is very similar to the recent work of \cite{Mauricio:2007vz}.  In this model the initial parton distribution is taken as thermal with the temperature chosen in order to reproduce the thermal dilepton number given by the hydrodynamic simulation.  Starting with the thermal initial condition at $\tau=0.2$ fm/c the total dilepton yield is found by integrating the free streaming result \ref{eq:A10} up to a final time of $\tau=1.0$ fm/c.  The details of this calculation is given in the appendix.  We now discuss the results.

We consider two scenarios.  The first is running the hydrodynamic simulation starting at $\tau_0=0.2$ fm/c until $T_c$.  The second scenario runs the free streaming model from $0.2\leq \tau \text{(fm/c)} \leq 1.0$.  Then at $1.0$ fm/c the hydrodynamic evolution is started and ran until $T_c$.  We should stress that the second model is not very realistic since the free streaming model is not asymptotic with the hydrodynamic evolution at $\tau=1$ fm/c.  A future work might use a more realistic model for the evolution of the distribution function then the proposed free streaming case.  For example, one could start with an initially anisotropic distribution which evolves to its thermal form from multi-quark scattering \cite{Xu:2005wj}, at which point a hydrodynamic evolution is started.  However, we expect the true result to lie between the two scenarios used in this work.

\begin{figure}
\includegraphics[width=0.45\textwidth]{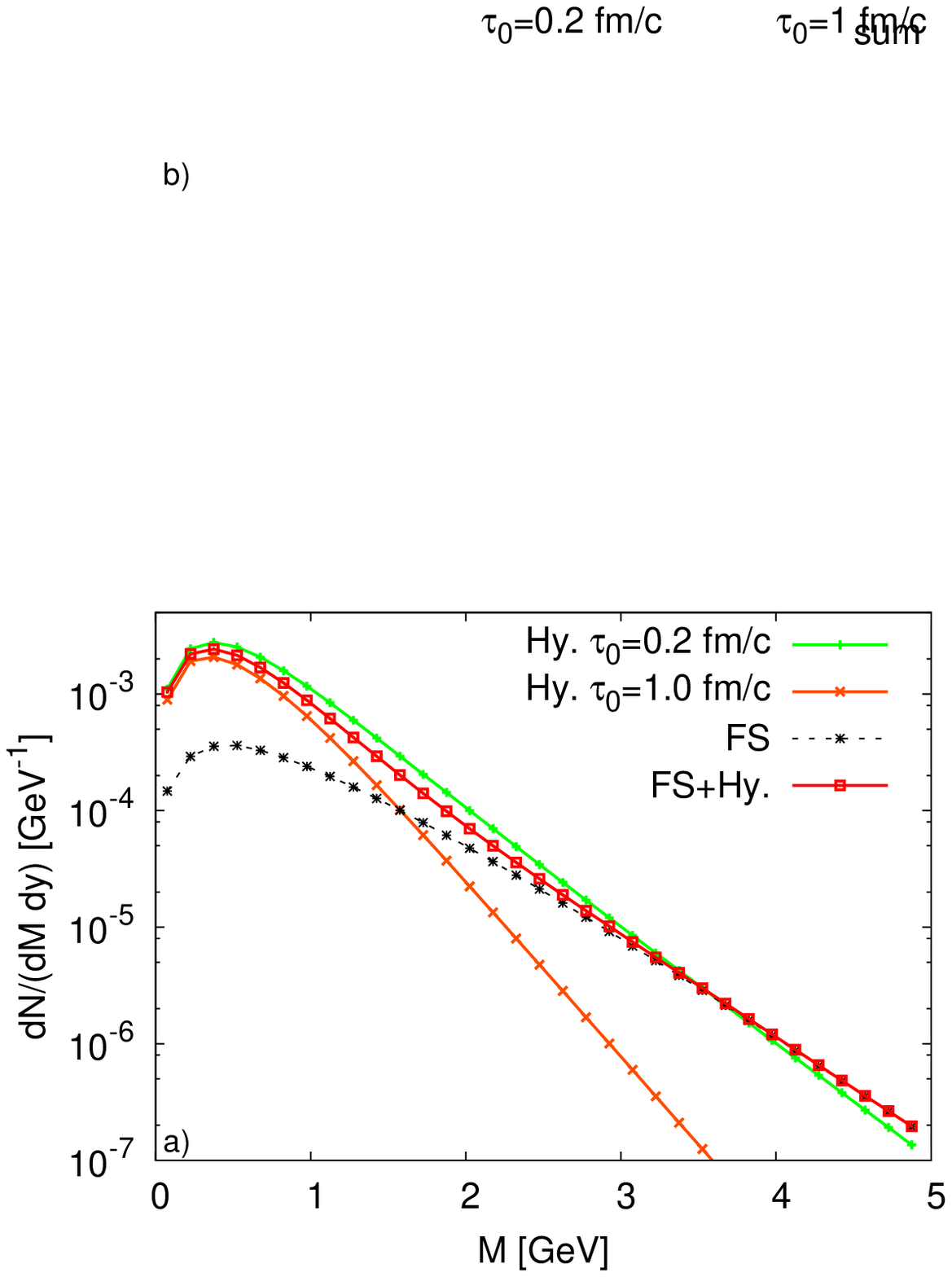}
\includegraphics[width=0.45\textwidth]{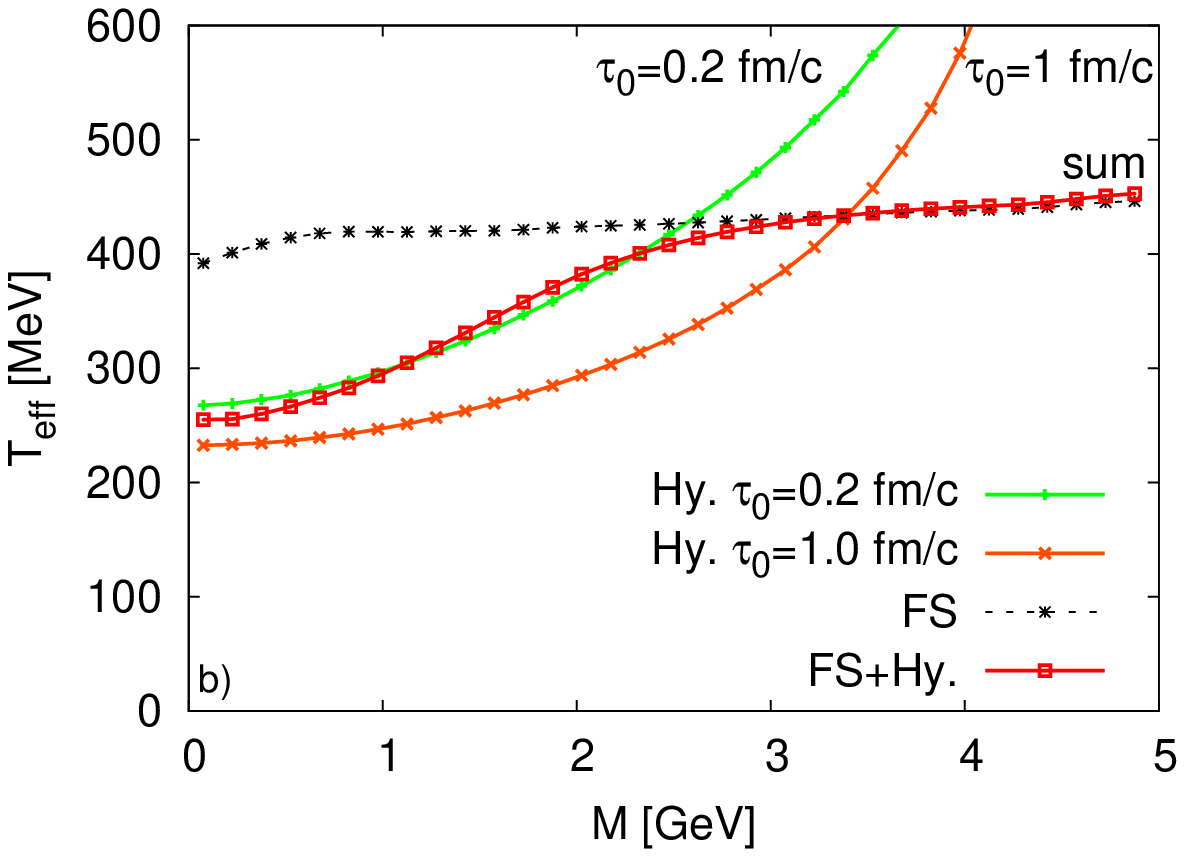}
\caption{\label{fig:summary}(color online) Left: Dilepton invariant mass spectra. Right: Effective temperature as a function of invariant mass.}
\end{figure}

Figure~\ref{fig:summary} shows both the invariant mass spectrum (left) and effective temperature (right) for the two scenarios outlined above. The curves to compare are the hydrodynamic simulation started at $\tau_0=0.2$ (labeled {\em Hy. $\tau_0=0.2$ fm/c}) and the sum of the hydrodynamic simulation started at $\tau_0=1$ and the free-streaming (labeled {\em FS+Hy.}).  We first note that the invariant mass spectrum is qualitatively the same for the two scenarios and it would not be possible to discern between the two scenarios from experimental data.  Qualitative differences do appear in the $T_{eff}$ spectrum.  First we find that the free-streaming with hydro solution mimics the early time hydro only solution at low masses.  However, at high masses the two result diverge when the viscous correction can no longer be trusted.  While the early hydrodynamic solution increases greatly with mass the free streaming solution flattens off at higher mass.  We therefore argue that through a detailed analysis of $q_\perp$ spectra one could hopefully extract information on the thermalization time, viscosity to entropy ratio and thermalization mechanism in heavy-ion collisions.

\section{Conclusions}

In conclusion, we have calculated the first viscous correction to dilepton production from leading order q\={q} annihilation.  The rates are then integrated over the space-time history of a viscous hydrodynamic simulation of RHIC collisions.  We argue that a thermal description is only reliable for invariant masses less than $\approx(2 \tau_0 T_0^2)/(\eta/s)$ and above this a kinetic description is required.  For $\eta/s=0.2$ and $\tau_0=1$ fm/c this corresponds to $M\lsim 4.5$ GeV.  We have shown that viscosity does not change the invariant mass distribution but strongly modifies the transverse momentum distribution and can therefore be used to extract information on both the viscosity to entropy ratio as well as the thermalization time.  Finally, we have also made comparisons with an initially free streaming QGP.  Qualitative differences in transverse momentum are seen, which could again possibly be used to learn about the thermalization mechanism.

{\bf Acknowledgments.}
\\
The authors are indebted to Ulrich Heinz for pointing out a critical mistake in a previous version of the manuscript.  We would like to thank Derek Teaney and Clint Young for useful discussions.  This work is partially supported by the US-DOE grants DE-FG02-88ER40388 and DE-FG03-97ER4014. 

\appendix
\section{Dilepton yields from a free streaming QGP}

In this appendix the dilepton yields are derived for a free streaming boost invariant expansion.  The starting point is the collision-less Boltzmann equation

\begin{equation}\label{eq:bz}
p^{\mu}\partial_{\mu}f(p,x)=0\,,
\end{equation}

where $f(p,x)$ will be considered as the phase-space distribution for the quark and anti-quark.  Under the assumption of boost invariance as well as homogeneity in the transverse plane the Boltzmann equation can be written as

\begin{equation}
\partial_\tau f-\frac{\tanh\chi}{\tau}\partial_\chi f=0\,,
\label{eq:fs}
\end{equation}

where $\chi=y-\eta_s$.  The initial condition is such that the quark distribution is isotropic and starts from local thermal equilibrium, $f(p,\tau=\tau_0)=\frac{1}{e^{p_0/T}+1}$.  One can write $p_0=u\cdot p=p_\perp \cosh(\chi)$ by using the assumption of boost invariance and homogeneity in the transverse plane.  We note that even for quarks out of equilibrium it still might be useful to use the equilibrium form of the distribution function where $T$ is instead considered as an effective temperature describing the initial state.  The solution of eq.~\ref{eq:fs} at any time $\tau$ is

\begin{equation}
f(p,x)=\frac{1}{e^{\frac{p_{\perp}}{T}\sqrt{1+\sinh^2(\chi)\left(\frac{\tau}{\tau_0}\right)^2}}+1}\,.
\end{equation}

With the explicit form of the distribution function available one can calculate the dilepton rates using the same kinetic theory expression used before (see eq.~\ref{eq:KT})

\begin{equation}
\frac{dN}{d^4q}=\int d^4x\int\frac{d^3p_1}{(2\pi)^3}\frac{d^3p_2}{(2\pi)^3}
f(p_1,x)f(p_2,x)v_{12}\sigma\delta^{(4)}(p_1+p_2-q)\,.
\end{equation}

Making use of the expressions for the relative velocity and cross section as quoted earlier the above equation can be expressed in the following form

\begin{equation}
\frac{dN}{d^4q}=B\int d^4x\int d^4p_1d^4p_2\delta(p_1^2)\delta(p_2^2)\delta^{(4)}(p_1+p_2-q)f(p_1,x)f(p_2,x)\,,
\end{equation}

where $B=\frac{32\pi\alpha^2 e_q^2}{(2\pi)^6}$.

First let us quote some well known identities:

\begin{eqnarray}
d^4x&=&\tau d\tau d\eta_s d^2x_{\perp}=\pi R^2\tau d\tau d\eta_s \nonumber\\
d^4p&=&\frac{1}{2}dp^2dy_pp_{\perp}dp_{\perp}d\phi_p \nonumber\\
d^4q&=&M dM q_{\perp}dq_{\perp} dy d\phi \nonumber\\
\delta^{(4)}(P-q)&=&4\delta(P^2-M^2)\delta(y_p-y)\delta(P_{\perp}-q_{\perp})\delta(\phi_P-\phi) \nonumber
\end{eqnarray}

where $P^{\mu}=p_1^{\mu}+p_2^{\mu}$ and $y$ and $\phi$ are rapidity and angle in the 
transverse plane.  We place a subscript $P$ on quantities to indicate they are derived from $P^{\mu}$.  The free streaming dilepton rate can now be expressed as

\begin{eqnarray}
\frac{dN}{d^4q}= B\int d^4x
\int dy_1p_{1,\perp}dp_{1,\perp}d\phi_1dy_2p_{2,\perp}dp_{2,\perp}d\phi_2f(p_1,x)f(p_2,x)
\times \nonumber\\
\delta(P^2-M^2)\delta(P_{\perp}^2-q_{\perp}^2)\delta(y_P-y)\delta(\phi_p-\phi)
\end{eqnarray}

Since the distribution function is boost invariant the integral over $\eta_s$ is trivial due to the delta function, $\delta(y_P-y)$.  By defining $y_{\pm}=y_1\pm y_2$ and $\phi_{\pm}=\phi_1\pm\phi_2$ the delta functions can be rewritten as

\begin{eqnarray}
\delta(P^2-M^2)&=&\frac{1}{2p_{1,\perp}p_{2,\perp}}\delta(\cosh y_--\cos\phi_--\frac{M^2}{2p_{1,\perp}p_{2,\perp}}) \nonumber \\
\delta(P_{\perp}^2-q_{\perp}^2)&=&\frac{1}{2p_{1,\perp}p_{2,\perp}}\delta(\cos\phi_-+\frac{p_{1,\perp}^2+p_{2,\perp}^2-q_{\perp}^2}{2p_{1,\perp}p_{2,\perp}}) \nonumber
\end{eqnarray}

After rewriting the integration variables as $dy_1dy_2=\frac{dy_+dy_-}{2}$ and $d\phi_1d\phi_2=\frac{d\phi_+d\phi_-}{2}$ the integral over $\phi_-$ and $y_-$ can be done explicitly yielding 

\begin{equation}
\frac{dN}{MdMdyq_{\perp}dq_{\perp}}=4\pi^2R^2 B\int\tau d\tau
\int dy_+ \frac{p_{1,\perp}dp_{1,\perp}p_{2,\perp}dp_{2,\perp}}{(2p_{1,\perp}p_{2,\perp})^2}\frac{1}{\lvert\sin\phi_-\rvert}\frac{1}{\sinh y_-}f(p_1,x)f(p_2,x) 
\end{equation}

where 

\begin{eqnarray}
\lvert\sin\phi_-\rvert&=&\frac{\sqrt{((p_{1,\perp}+p_{2,\perp})^2-q_{\perp}^2)
(q_{\perp}^2-(p_{1,\perp}-p_{2,\perp})^2)}}{2p_{1,\perp}p_{2,\perp}} \nonumber\\
\sinh y_-&=&\frac{\sqrt{(M^2+q_{\perp}^2-(p_{1,\perp}+p_{2,\perp})^2)(M^2+q_{\perp}^2-(p_{1,\perp}-p_{2,\perp})^2)}}{2p_{1,\perp}p_{2,\perp}}
\end{eqnarray}

The delta function in the above equation enforces the following constraints

\begin{eqnarray}\label{eq:constraint}
\left\lvert\frac{p_{1,\perp}^2+p_{2,\perp}^2-q_{\perp}^2}{2p_{1,\perp}p_{2,\perp}}\right\rvert&\leqslant& 1 \nonumber \\
\frac{M^2+q_{\perp}^2-p_{1,\perp}^2-p_{2,\perp}^2}{2p_{1,\perp}p_{2,\perp}}&\geqslant& 1
\end{eqnarray}

Let us make a further shift of variables, $p_{\pm}=p_{1,\perp}\pm p_{2,\perp}$. The
constraints (\ref{eq:constraint}) then take particularly simple form.  The final expression is

\begin{eqnarray}\label{eq:fsrate}
&&\frac{dN}{dM^2dydq_{\perp}^2}=\pi R^2 \frac{N_c \alpha^2 e_q^2}{48\pi^4} \int\tau d\tau
\int_{-\infty}^{+\infty} dy_+\int_{q_{\perp}}^{\sqrt{M^2+q_{\perp}^2}}dp_+\int_{-q_{\perp}}^{q_{\perp}}dp_-(p_+^2-p_-^2) \times \nonumber\\
&&\frac{1}{\sqrt{(M^2+q_{\perp}^2-p_+^2)(M^2+q_{\perp}^2-p_-^2)}}\frac{1}{\sqrt{(p_+^2-q_{\perp}^2)(q_{\perp}^2-p_-^2)}}f\left(p_1,\tau\right)f\left(p_2,\tau\right)
\label{eq:A10}
\end{eqnarray}

where

\begin{eqnarray}
f\left(p_1,\tau\right)&=&\left[ 1+\exp\left(\frac{p_++p_-}{2T}\sqrt{1+\left(\frac{\tau}{\tau_0}\right)^2\sinh\left(\frac{y_++y_-}{2}\right)}\right)\right]^{-1}\nonumber\\
f\left(p_2,\tau\right)&=&\left[ 1+\exp\left(\frac{p_+-p_-}{2T}\sqrt{1+\left(\frac{\tau}{\tau_0}\right)^2\sinh\left(\frac{y_+-y_-}{2}\right)}\right)\right]^{-1}\nonumber\\
y_-&=&\sinh^{-1}\left[\frac{2\sqrt{\left(M^2+q_\perp^2-p_+^2\right)\left(M^2+q_\perp^2-p_-^2\right)}}{\left(p_+^2-p_-^2\right)} \right]
\end{eqnarray}


\begin{thebibliography}{MM}
\bibitem{Ollitrault:2006va}
  J.~Y.~Ollitrault,
  Pramana {\bf 67}, 899 (2006).

\bibitem{Teaney:2001av}
D.~Teaney, J.~Lauret and E.~V.~Shuryak,
arXiv:nucl-th/0110037. {\it ibid},
Phys.\ Rev.\ Lett.\  {\bf 86}, 4783 (2001)

\bibitem{Kolb:2003dz}
  P.~F.~Kolb and U.~W.~Heinz,
  arXiv:nucl-th/0305084.

\bibitem{Dusling:2007gi}
  K.~Dusling and D.~Teaney,
  arXiv:0710.5932 [nucl-th].
\bibitem{Song:2007ux}
  H.~Song and U.~W.~Heinz,
  arXiv:0712.3715 [nucl-th].
\bibitem{Baier:2006gy}
  R.~Baier and P.~Romatschke,
  Eur.\ Phys.\ J.\  C {\bf 51}, 677 (2007)
  [arXiv:nucl-th/0610108].

\bibitem{Romatschke:2007jx}
  P.~Romatschke,
  Eur.\ Phys.\ J.\  C {\bf 52}, 203 (2007)
  [arXiv:nucl-th/0701032].

\bibitem{Bozek:2007qt}
  P.~Bozek,
  arXiv:0712.3498 [nucl-th].
\bibitem{Song:2007fn}
  H.~Song and U.~W.~Heinz,
  arXiv:0709.0742 [nucl-th].
\bibitem{Chaudhuri:2007qp}
  A.~K.~Chaudhuri,
  arXiv:0708.1252 [nucl-th].

\bibitem{WongBk}
C.-Y.~Wong, {\em Introduction to High-Energy Heavy-Ion Collisions}, (World Scientific, 1994).
\bibitem{Gelis:2001xt}
  F.~Gelis, D.~Schiff and J.~Serreau,
  Phys.\ Rev.\  D {\bf 64}, 056006 (2001)
  [arXiv:hep-ph/0104075].

\bibitem{Teaney:2003kp}
  D.~Teaney,
  Phys.\ Rev.\  C {\bf 68}, 034913 (2003)
  [arXiv:nucl-th/0301099].

\bibitem{Arnold:2000dr}
  P.~Arnold, G.~D.~Moore and L.~G.~Yaffe,
  JHEP {\bf 0011}, 001 (2000)
  [arXiv:hep-ph/0010177].

\bibitem{GrootBk}
S. de Groot, W. van Leeuven and Ch. van Veert, {\em Relativistic Kinetic Theory} (North-Holland, Amsterdam, 1980).
\bibitem{OG}
M. Grmela, H.C. \"{O}ttinger, Phys. Rev. E {\bf 56}, 6620 (1997). H.C. \"{O}ttinger, M. Grmela, Phys. Rev. E {\bf 56}, 6633 (1997). H.C. \"{O}ttinger, Phys. Rev. E {\bf 57}, 1416 (1993).
\bibitem{Gyulassy:1997ib}
  M.~Gyulassy, Y.~Pang and B.~Zhang,
  Nucl.\ Phys.\  A {\bf 626}, 999 (1997)
  [arXiv:nucl-th/9709025].
\bibitem{Kapusta:1992uy}
  J.~I.~Kapusta, L.~D.~McLerran and D.~Kumar Srivastava,
  Phys.\ Lett.\  B {\bf 283}, 145 (1992).
\bibitem{Mauricio:2007vz}
  M.~Martinez and M.~Strickland,
  arXiv:0709.3576 [hep-ph].
\bibitem{Xu:2005wj}
  X.~M.~Xu, P.~Ru, H.~J.~Weber and H.~J.~Weber,
  Phys.\ Lett.\  B {\bf 629}, 68 (2005)
  [arXiv:hep-ph/0509046].

\end{thebibliography}
\end{document}